\documentclass{emulateapj}
\def\Msun{$M_\odot$}

\def\mloss{$\overline{\dot M}_{\rm loss}$}
\def\mcore{$\overline{\dot M}_{\rm core}$}
\def\kms{km$\cdot$s$^{-1}$}

\defcitealias{bot09}{B09}
\defcitealias{val09}{V09}
\defcitealias{thesis}{P06}
\defcitealias{s07}{S07}
\begin{document} 

\title{EC-SNe from super-AGB progenitors: theoretical models vs. observations}

\author{M.~L. Pumo\altaffilmark{1,2}, M. Turatto\altaffilmark{2}, M.~T. Botticella\altaffilmark{3}, A. Pastorello\altaffilmark{3}, S. Valenti\altaffilmark{3}, L. Zampieri\altaffilmark{1}, S. Benetti\altaffilmark{1}, E. Cappellaro\altaffilmark{1}, F. Patat\altaffilmark{4}} 
\shortauthors{Pumo et al.}
\affil{\altaffilmark{1}INAF - Osservatorio Astronomico di Padova, Vicolo dell'Osservatorio 5, I-35122 Padova, Italy} 
\affil{\altaffilmark{2}INAF - Osservatorio Astrofisico di Catania, via S. Sofia 78, I-95123 Catania, Italy}
\affil{\altaffilmark{3}Astrophysics Research Centre, School of Mathematics and Physics, Queen's University Belfast, Belfast BT7 1NN, UK}
\affil{\altaffilmark{4}European Southern Observatory, K. Schwarzschild Str. 2, 85748 Garching b. M\"unchen, Germany}
\date{Received ... ; accepted ...}

\begin{abstract} 
Using a parametric approach, we determine the configuration of super-AGB 
stars at the explosion as a function of the initial mass and metallicity, in order to verify 
if the EC-SN scenario involving a super-AGB star is compatible with the observations regarding 
SN2008ha and SN2008S. The results show that both the SNe can be explained in terms of EC-SNe 
from super-AGB progenitors having a different configuration at the collapse. The impact of 
these results on the interpretation of other sub-luminous SNe is also discussed.
\end{abstract}


\section{Introduction}
\label{sec:intro}

It is widely accepted that there are two main explosion mechanisms leading to supernova (SN) 
events \citep[e.g.][]{woosley86,hille00}: the thermonuclear runaway in white dwarfs 
approaching the Chandrasekhar mass and the core collapse of stars with initial mass $\gtrsim 
8$\Msun (CC-SNe). From an observational point of view, the former mechanism originates the 
relatively homogeneous type Ia SNe. The latter produces a huge variety of displays (different 
energetics and amounts of ejected material, reflecting on heterogeneous light curves and 
spectra evolutions), which are thought to be linked to the possible interaction of the CC-SN 
ejecta with circumstellar material (CSM) and the different configuration of the CC-SN 
progenitor at the time of the explosion \citep*[e.g.][]{filip97,hamuy03,turatto03,turatto07}.

However the exact nature of CC-SN progenitors (initial mass, stellar structure and composition 
at the explosion) and the type of collapse (iron core collapse or neon-oxygen core collapse 
triggered by electron captures) are far from being well-established. There are still 
ambiguities that arise, on the theoretical side, from the uncertainties in modelling stellar 
evolution and the explosion mechanism \citep[e.g.][]{woosley02,heger03} and, on the observational 
side, from the sparse direct detections of progenitor stars and a controversial classification 
of some events \citep[e.g.][]{smartt08,turatto07}.

A number of faint transients have been recently discovered whose nature is still ambiguous and 
extensively debated. In particular SN~2008S received a SN designation by \citet[][]{stan08}, 
but \citet[][]{steele08} considered it as a SN ``impostor'', and \citet[][]{smith08} as the 
exotic eruption of a luminous blue variable (LBV) object with a relatively low-mass, highly 
obscured progenitor ($\lesssim 15$\Msun). An eruptive origin was invoked also for two other similar 
transients \citep[M 85 OT2006-1 and NGC 300 OT2008-1;][]{kulk07,Berger09,bond09}. However, work based 
on multi-wavelength follow-up of the transients and mid-IR images analysis of the pre-explosion 
environments not only did not rule out a SN origin \citep[][]{pastorello07,Prieto08} but even suggested 
that they may be CC-SNe triggered by electron-capture reactions (so-called EC-SNe) involving a 
super-asymptotic giant branch (super-AGB) star \citep[][]{thompson08}. The long-term multiwavelength 
monitoring of the SN2008S and new comparisons with the two aforementioned transients seem to support the 
EC-SN interpretation \citep[][B09 hereafter]{bot09}. In particular B09 favor a scenario where the SN2008S 
progenitor is a super-AGB star embedded in an optically thick circumstellar shell. This 
conclusion is based on (1) the fact that the pre-explosion luminosity of the progenitor is in 
plausible agreement with the super-AGB models, (2) the capability of super-AGB stars to form 
thick circumstellar shells through mass-loss during the thermal pulses phase, (3) the 
similarity in the total radiated energy of SN2008S with that of other faint SNe, (4) the 
moderate velocities ($\sim 3000$\kms) of the ejecta, and (5) a low but significant 
mass ($0.0014 \pm 0.0003$\Msun) of ejected $^{56}$Ni.

An EC-SN explanation has been suggested also for SN2008ha \citep[][V09 hereafter]{val09}, 
although an iron CC-SN involving a massive star (initial mass $\gtrsim 25-30\,M_{\odot}$) plus 
fallback onto the collapsed remnant can not be ruled out. At first this object was included among 
the SN2002cx-like variety of peculiar type Ia SNe \citep[][]{Li03,jha06,phillips07}, but V09 
reviewed the thermonuclear scenario on the basis of the photometric and spectroscopic 
similarity to low-luminosity CC-SNe, concluding that all SN 2002cx-like objects could be 
indeed faint, stripped-envelope CC-SNe and that SN2008ha represents the faint tail in the 
luminosity distribution of this SN family. However, \citet{foley09} did not definitively rule 
out the thermonuclear origin of the SN2002cx-like objects, and proposed to explain the 
peculiarity of SN2008ha in terms of an ``accretion-induced'' collapse \citep[so-called AIC 
mechanism; see][for details]{metzger09}.

So far no clear picture has emerged and different scenarios may explain the aforementioned 
faint transients, especially because a detailed scrutiny of the super-AGB progenitors 
configuration at the explosion, which is crucial for a comparison with SN observables, 
is still missing. In the light of the most recent super-AGB stars models (e.g. \citealt{sp06}, 
SP06 hereafter; \citealt{thesis}, P06 hereafter; \citealt{s07}, S07 hereafter; \citealt{Poel08}), 
in this Letter we discuss in detail the possible link between EC-SNe from super-AGB progenitors and 
these transients using a parametric approach. After a brief sketch of the 
super-AGB stars evolution, we determine their configuration at the explosion as a function of 
the initial mass and metallicity from the most recent grids of super-AGB stellar models, and 
then we investigate if such a configuration is compatible with the observations.


\section{Evolution and final fate of super-AGB stars}
\label{sec:sagb}

Current view on stellar evolution reveals the existence of two critical initial masses 
$M_{up}$, defined as the minimum initial mass above which C-burning ignites, and $M_{mas}$, 
corresponding to the minimum initial mass for the completion of all the nuclear burning phases 
leading to an iron core collapse \citep[e.g.][]{woosley02}. So-called super-AGB stars fill the 
gap between $M_{up}$ and $M_{mas}$. After the central He-burning, these stars develop partially 
degenerate CO cores, which are massive enough to exceed the threshold value for C-burning 
ignition, that transforms the core into a degenerate NeO mixture (e.g. P06; S07). The gravothermal 
energy released by the core contraction after the central He-exhaustion induces the occurrence 
of the second dredge-up or dredge-out phenomenon \citep[e.g.][]{ritosa99}, which increases the He 
abundance in the envelope up to $Y \sim 0.38$ (see e.g. Fig. 1 in \citealt{pumo08}, but also SP06 and S07).

When C-burning stops in the core, the physical conditions are not suitable for the 
Ne-ignition and the structure of super-AGB stars becomes very similar to that of AGB stars, 
having an inert core surrounded by a He- and H-burning shell. Thus super-AGBs can be 
considered high-mass equivalent to AGBs and, as such, they may suffer thermal pulses and lose 
mass as ``normal'' (but quite massive and luminous) AGB stars \citep[e.g.][and references 
therein]{pumo08}. In this situation the H-free core grows in mass and, if it reaches the 
Chandrasekhar limit ($M_{CH} \sim 1.37 M_{\odot}$; \citealt{n84}), EC reactions are activated 
first on $^{24}$Mg and $^{24}$Na and then on $^{20}$Ne and $^{20}$F. Since these reactions have 
the effect to release entropy and decrease the electron mole number $Y_e$, O-ignition and core 
collapse are induced almost simultaneously, and a deflagration front (incinerating the material 
into a nuclear statistical equilibrium composition) forms when the central density reaches 
$2.5\cdot10^{10}$ g$\cdot$cm$^{-3}$ \citep[e.g.][]{Hillebrandt84}. However the O-deflagration is too ``weak'' 
to avoid the core collapse, so it proceeds up to neutron star density \citep[see][for details]{Miyaji80,n84}. 
This mechanism, leaving a neutron star as remnant, presents distinctive features \citep[e.g.][]{kitaura06,wanajo09}: 
low explosion energy ($\sim 10^{50}$erg), large Ni/Fe ratio ($\simeq1$-$2$) and ejection of small 
amount of $^{56}$Ni (between $\sim 0.002$ and $\sim0.004$\Msun).

Whether or not the stellar core reaches the threshold value $M_{CH}$ for triggering the EC-SN, 
depends on the interplay between mass loss and core growth \citep[e.g.][]{woosley02,h05}. If 
mass loss is high enough, the envelope is lost before the core can reach $M_{CH}$, and the endpoint 
of super-AGB evolution is a NeO white dwarf (WD). On the contrary, if the mass loss is not so 
efficient, the super-AGB star evolves into an EC-SN. The critical initial mass setting the transition 
between super-AGBs that evolve into a NeO WD and the ones that undergo EC-SN is referred to as $M_{N}$.

\begin{figure}
\epsscale{.90}
\plotone{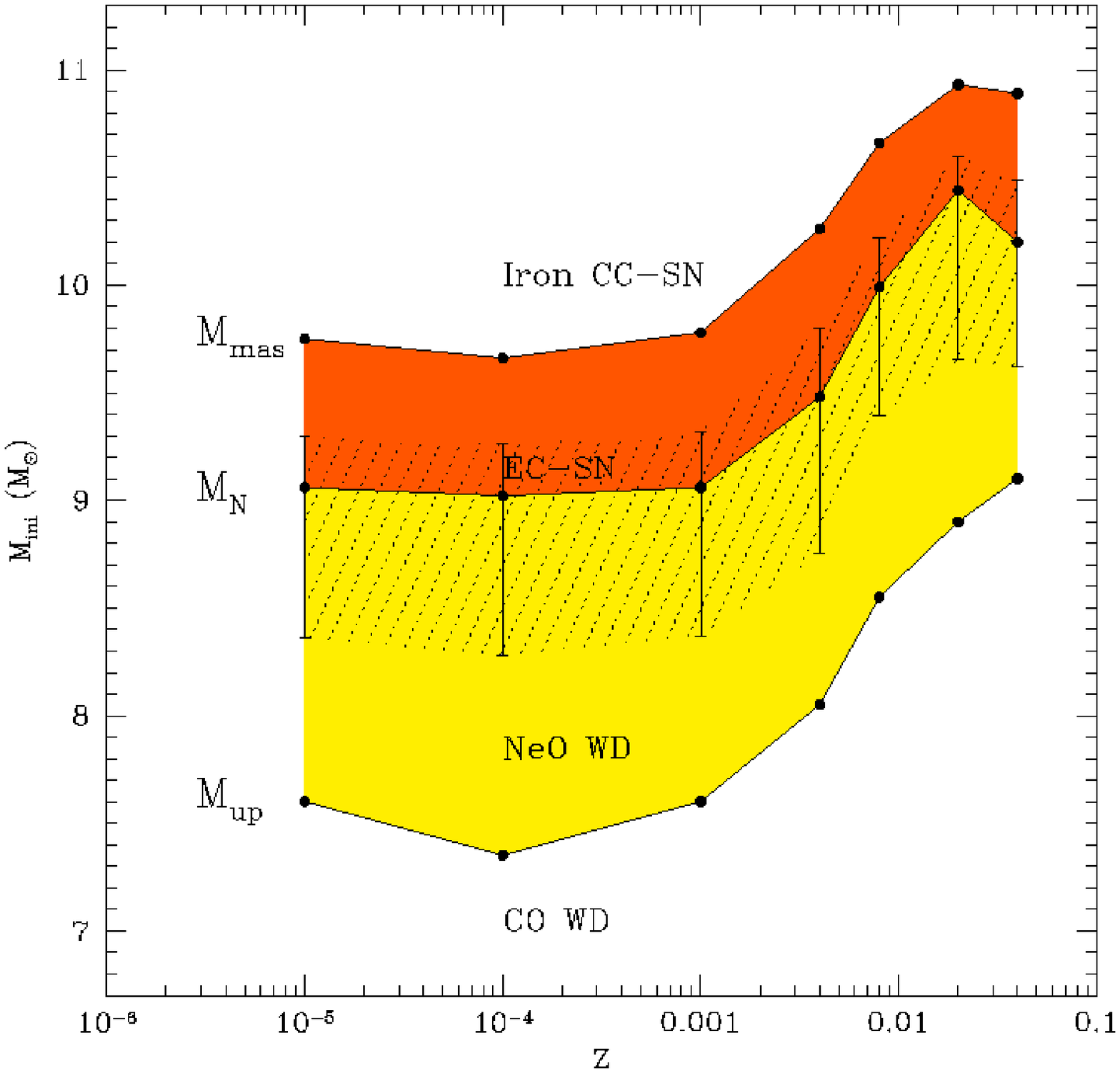}
\caption{Mass transitions $M_{up}$, $M_{N}$ and $M_{mas}$ as a function of the initial metallicity $Z$. 
The error bars indicate the possible range of variation of $M_{N}$ (the dashed area is used to mark 
the range of uncertainty) caused by indeterminations on the mass loss and core growth rates. The different 
outcomes of stellar evolution are also reported: CO WD for stars having initial mass less than $M_{up}$, 
NeO WD for stars having initial mass between $M_{up}$ and $M_{N}$ (see the yellow zone), EC-SN for stars 
having initial mass between $M_{N}$ and $M_{mas}$ (see the orange zone), iron CC-SN for stars having initial 
mass greater than $M_{mas}$. (Figure adapted from \citet[][]{pumo_tacchini07}. Details on the data can be 
found in \citetalias{thesis} and \citetalias{s07}).\label{fig1}}
\end{figure}

Recent studies (SP06; P06; S07; \citealt{Poel08}) have shown that EC-SN channel may exist, but 
uncertainties in mass loss and core growth rates hamper any conclusions on the exact fraction of super-AGBs 
evolving into this channel (see Fig.~\ref{fig1}). So the actual realization of the EC-SN mechanism
in super-AGBs should be taken with caution. Nevertheless, it is fair to consider this 
scenario and its implications.


\section{Outcome of EC-SNe from super-AGB progenitors.}
\label{sec:outcome}

As already mentioned in Sect.~\ref{sec:intro}, for the comparison with SN observations it is 
crucial to know the configurations of the progenitors at the explosion. In fact, as explained below, 
the least massive super-AGB progenitors (i.e. super-AGBs with initial mass close to $M_{N}$) may have 
lost almost all their envelope at the time of the explosion, while the most massive ones (i.e. 
super-AGBs with initial mass close to $M_{mas}$) may still retain a massive ($\sim 
8$-$9$\Msun) envelopes. Also the CSM can be different with dense shells in 
proximity of the most massive super-AGBs progenitors and much looser CSM in proximity of the 
lower-mass progenitors.

These diversities imply that EC-SNe may be observed as relatively faint Type II SNe (IIP or 
IIL depending on the mass of the H-rich envelope) with presumably low degree of CSM interaction, 
as Type IIb SNe having stronger interaction with dense, structured and possibly He-enhanced 
(thanks to the second dredge-up or dredge-out) CSM, up to stripped envelope SNe.

In Tab.~\ref{tabl1} we summarise the main parameters describing the structure of super-AGB 
stars of different initial mass and metallicity at the time of explosion. These were 
built starting with the calculation of the stellar structure at the beginning of the thermally 
pulsing super-AGB (TP-SAGB) phase from the grids of fully super-AGB stellar models reported 
in \citetalias{thesis} and \citetalias{s07}. Afterwards, the structure at the explosion was 
calculated, considering that the envelope mass at the explosion may be estimated as follows:
\begin{equation}
M_{env}^{EC-SN}=M_{\star}^{postCB}-M_{CH}-\Delta M_{loss}^{postCB} \, ,
\end{equation}
where $M_{\star}^{postCB}$ is the total stellar mass at the beginning of the TP-SAGB phase 
and $\Delta M_{loss}^{postCB}$ is the mass lost during the TP-SAGB 
evolution. This last term can be estimated from the relation: 
\begin{equation}
\Delta M_{loss}^{postCB} = \overline{\dot M}_{loss} \cdot \Delta t_{CH} \, ,
\end{equation}
where \mloss\, is the averaged mass loss rate 
during the TP-SAGB evolution and $\Delta t_{CH}$ is the time interval from the beginning of 
the TP-SAGB phase until core mass reaches $M_{CH}$, given by 
\begin{equation}
\Delta t_{CH}=\frac{M_{CH}-M_{core}^{postCB}}{\overline{\dot M}_{core}} \, .
\end{equation}
In this expression, $M_{core}^{postCB}$ is equal to the H-free core mass at the beginning of the
TP-SAGB phase and \mcore\, is equal to the averaged core growth rate during the TP-SAGB evolution.

The values reported in Tab.~\ref{tabl1} are calculated considering a typical core growth rate
of \mcore=$5\cdot10^{-7}$\Msun\,$yr^{-1}$ \citep[e.g.][]{Poel06,Poel08} and choosing a 
reasonable value of $\zeta\equiv \frac{\overline{\dot M}_{\rm loss}}{\overline{\dot M}_{\rm 
core}}=70$ during the TP-SAGB evolution (``realistic'' values for this ratio 
vary from $\sim 35$ to $\sim 100$; see S07, for details). This choice for the $\zeta$ value 
corresponds to \mloss$=3.5\cdot10^{-5}$\Msun\,$yr^{-1}$ and is consistent with 
the value deduced from the observations (\citealt{Prieto08} estimated a mass loss rate 
$\gtrsim10^{-5}$\Msun\,$yr^{-1}$ for the progenitor of the SN2008S). 

In the two last columns of Tab.~\ref{tabl1} we report the total ejected mass evaluated 
assuming a mass cut of $\sim$ 1.36\Msun \citep[e.g.][]{kitaura06}, and the maximum distance 
travelled by the CSM lost during the TP-SAGB evolution, calculated assuming an average 
wind velocity of 10 \kms\,.

Although this parametric approach to determinate the structure of super-AGB stars is 
simplistic, it is completely consistent with the approach used to determine the fraction of 
super-AGB stars evolving into EC-SNe by P06 and S07, whose models are the basis for our 
calculation. In addition it should be noted that more sophisticated synthetic models for 
super-AGB stars cannot presently reach a much higher precision because no stellar models 
describing the TP-SAGB evolutionary phase are available at the moment.

\begin{table}[ht]
  \begin{center}
  \caption{\label{tabl1}}
  \begin{tabular}{c|ccccc}
  M$_{\star}^{ini}$ & M$_{\star}^{postCB}$ & M$_{core}^{postCB}$ & $\Delta t_{CH}$ & M$_{ej}$    & D$_{CSM}^{max}$  \\
  $[$\Msun$]$       & $[$\Msun$]$          & $[$\Msun$]$         & $[10^{5}\,yr]$  & $[$\Msun$]$ & $[10^{5}\,A.U.]$ \\
  \tableline
  & \multicolumn{5}{|c}{Z=0.004}\\
  \tableline
   9.48 &  9.36 & 1.24 & 3.02 & $\sim$ 0.01 & 6.4  \\
   9.55 &  9.43 & 1.26 & 2.24 & 0.3         & 6.0  \\
   9.75 &  9.63 & 1.27 & 2.04 & 1.1         & 4.3  \\
   9.95 &  9.84 & 1.31 & 1.25 & 4.1         & 2.6  \\
  10.15 & 10.04 & 1.35 & 0.46 & 7.0         & 1.0  \\
  10.25 & 10.14 & 1.37 & 0.07 & 8.5         & 0.1  \\
  \tableline 
  & \multicolumn{5}{|c}{Z=0.008}\\
  \tableline 
   9.99 &  9.77 & 1.25 & 2.39 & $\sim$ 0.01 & 5.0  \\
  10.01 &  9.82 & 1.26 & 2.23 & 0.3         & 4.7  \\
  10.15 &  9.97 & 1.28 & 1.77 & 2.4         & 3.7  \\
  10.35 & 10.17 & 1.31 & 1.11 & 4.9         & 2.4  \\
  10.55 & 10.38 & 1.35 & 0.46 & 7.4         & 1.0  \\
  10.65 & 10.49 & 1.36 & 0.10 & 8.8         & 0.2  \\
  \tableline
  & \multicolumn{5}{|c}{Z=0.02}\\
  \tableline 
  10.44 & 10.15 & 1.24 & 2.52 & $\sim$ 0.01 & 5.3  \\
  10.46 & 10.16 & 1.25 & 2.47 & 0.2         & 5.2  \\
  10.55 & 10.25 & 1.27 & 2.02 & 1.8         & 4.3  \\
  10.65 & 10.35 & 1.29 & 1.52 & 3.7         & 3.2  \\
  10.75 & 10.45 & 1.32 & 1.01 & 5.5         & 2.1  \\
  10.85 & 10.55 & 1.34 & 0.51 & 7.4         & 1.1  \\
  10.92 & 10.63 & 1.36 & 0.10 & 8.9         & 0.2  \\
  \end{tabular}
  \tablecomments{Selected features (see text in Sect.~\ref{sec:outcome}) of the super-AGBs models as a function of the initial stellar mass for different $Z$ values. The first and the last row in each set of models with a given $Z$ refers to a super-AGB star with initial mass equal to $M_{N}$ and $M_{mas}$ (to less than $0.01$\Msun), respectively.}
\end{center}
\end{table}
%


\section{Discussion and Conclusions}

The two events SN2008ha and SN2008S find a reasonable interpretation in the aforementioned
scenario, and the progenitor mass to be associated with these SNe can be determined, considering the 
best ``global'' matching between the features of the super-AGBs models and the observed properties. 

Assuming an initial metallicity $Z$ for the progenitors from $\sim 0.008$ to $\sim 
0.02$ (see e.g. V09; B09; \citealt{foley09}, for details on the metallicity determination), 
one obtains that a super-AGB star with initial mass slightly above $M_{N}$ has M$_{core}^{postCB} 
\lesssim 1.25$-$1.26$\Msun\, (cf. second row in the sets of models in Tab.~\ref{tabl1}), while a 
super-AGB star with initial mass $\sim (M_{N}$+$0.5)$\Msun\, has M$_{core}^{postCB} \sim 
1.34$-$1.35$\Msun\, (cf. the row before the last in the sets of models in Tab.~\ref{tabl1}). As a 
consequence the time $\Delta t_{CH}$ necessary to the H-free core to reach $M_{CH}$ is $\gtrsim 
2.2$-$2.5\cdot10^{5}\,yr$ in the former case, and $\sim 5\cdot10^{4}\,yr$ in the latter one. 
This difference in the time elapsing between the beginning of TP-SAGB phase and 
the EC-SN event in the two cases, reflects on the configuration at the collapse. The super-AGB 
star with initial mass slightly above $M_{N}$ has time to expel almost all the envelope and, 
consequently, gives rise to a faint stripped-envelope SN characterised by a non H-rich\footnote{
We do not have accurate quantitative informations about the chemical composition of 
the ejecta (except for the $^{56}$Ni) to be compared to observations of SN2008ha, due to uncertainties
of both observational and theoretical nature. However, we speculate that the composition could be 
non H-rich. In fact, for this model the ejecta is composed by the H-free stellar layer between the mass cut 
and $M_{CH}$ (representing $\sim$ 5-15\% by mass of all the ejected mass) and by the remaining 
envelope mass at the explosion, whose ``initial'' H-rich composition can be deeply altered by 
the second dredge-up phenomenon, the so-called Hot Botton Burning, and the third dredge-up 
episodes.}
ejecta of $\lesssim 0.2$-$0.3$\Msun\, with no signatures of prompt CSM interaction, in agreement 
with the observations of SN2008ha (M$_{ej}$ in the range $0.1$-$0.5$\Msun\,, e.g. V09; \citealt{foley09}).
Assuming an average wind velocity of $10$\kms\,, $90$\% of the total expelled mass 
can be at a radial distance $\gtrsim 5 \cdot 10^{^4}A.U.$ when the EC-SN 
event takes place. The mean density of the CSM is expected to be $\lesssim5$cm$^{-3}$ 
(this value is likely to be even lower due to a decreased mass loss rate near the end
of the TP-SAGB phase when the mass of the envelope is significantly reduced), that 
could be sufficiently low not to give rise to significant interaction. The relatively 
low X-ray emission \citep[L$_{X} < 5 \cdot 10^{39}$ erg\,s$^{-1}$;][]{foley09} seems to 
support this idea, because the CSM can be an efficient X-ray radiator for much higher density 
\citep[$\sim 10^5$-$10^7$cm$^{-3}$;][]{Cheva01}. 

On the contrary, the super-AGB star with initial mass $\sim (M_{N}$+$0.5)$\Msun\, loses 
$\sim 1.6$-$1.8$\Msun\, in $\sim 5\cdot10^{4}\,yr$ --- consistently with the inferred 
duration of the so-called dust-enshrouded phase for SN2008S 
\citep[upper limit equal to $\sim 6\cdot 10^{4}\,yr$;][]{thompson08} --- 
and, besides maintaining a massive ($\sim 7.4$\Msun) envelope at the collapse, could be 
embedded within a thicker circumstellar envelope (mean density $\sim90$cm$^{-3}$). Observations 
of SN2008S \citepalias[][]{bot09} indicate the formation of a detached shell with an inner 
radius of $\sim 90\, A.U.$ and outer radius of $\sim 450\,A.U.$ having $\sim 0.08$\Msun\, of 
gas (van Loon, private communication). We could reproduce such a shell increasing the mass loss
rate by $\sim 15$ times above the average value for a relatively short period of $\sim 150\, yr$ 
as a consequence of a He-shell flash episode \citep[see][for details]{Matt07}. In addition, we 
find that $\sim 95$\% of the total expelled mass in the CSM is beyond the aforementioned detached 
shell, and these findings are in agreement with the presence of dust at a radial distance greater 
than $\sim 2 \cdot 10^{^3}A.U.$, as inferred from the MIR emission of SN2008S \citepalias[][]{bot09}. 

Thus the current understanding of super-AGB stellar evolution is quantitatively consistent 
with the available data on these two recent faint transients, that may be explained in terms 
of EC-SNe from super-AGB progenitors having a different configuration at the collapse, without 
resorting to ``exotic'' scenarios that are not free from uncertainties. As for the ``special'' 
eruption of LBV of relatively low mass proposed to explain the features of the SN2008S 
\citep[][]{smith08}, in addition to the problems for reconciling the ejecta velocity $\lesssim 
3000$\kms\, with a stellar eruption (B09), it is difficult to explain the fact that the slope 
of the late-time light curve of SN2008S \citep[but also that of the similar event NGC300-OT;][]{bond09} 
is surprisingly similar to that expected in a SN explosion when the main mechanism powering 
the SN luminosity is the radioactive decay of $^{56}$Co into $^{56}$Fe.
As for the AIC mechanism invoked for the SN2008ha, the main problem concerns the high 
velocity ($\sim 0.1$-$0.2 c$) not observed in the ejecta and the impossibility to synthesise 
the observed intermediate-mass nuclei, that are predicted by the ``standard'' (involving a 
single degenerate binary system) AIC model. The so-called ``enshrouded'' AIC model involving 
the merging of two WDs in a binary system \citep[][]{metzger09} might be somewhat less 
problematic. However the ejecta velocity, the amount of $^{56}$Ni and the production of 
intermediate-mass elements are still quantitatively poorly defined, and the role of the possible 
interaction between the disk wind and the outgoing SN shock has to be explored.

The weakness of the explosion and small amount of $^{56}$Ni synthesized make EC-SNe an obvious
explanation for low-luminosity core-collapse events with unusual properties that are related to
the pre-explosion mass loss episodes of their super-AGB progenitors and/or to the possible
ensuing ejecta/CSM interaction. However, it has been suggested that the EC channel may also 
account for the properties of some relatively ``normal'' type II SNe \citep[e.g.][]{CU00,kitaura06}, 
characterised by low luminosity, small amount of ejected $^{56}$Ni, extended plateaus (implying 
envelope mass of several \Msun) and slow expansion velocities \citep[e.g.][]{pastorello09}. 
To date, only for two objects of this class \citep[SN2005cs and SN2008bk;][]{mau06,li06,matt08} 
clear evidence has been found for low mass progenitors on pre-explosion images, but the fact 
that they are super-AGBs is strongly questionned \citep*[e.g.][]{eldridge07}. Thus, it remains 
to be seen what fraction (if any) of low luminosity type II SNe are EC-SNe and what other, instead, 
are more usual iron CC-SNe that experience less energetic than normal explosions \citep[as, for example, 
if some of them are sufficiently massive to undergo fallback onto the collapsed remnant; see e.g.][]{Zampieri03}.

The wide variety of displays expected for EC-SNe may be of interest also in understanding the two 
unusual events, SN2005E and SN2005cz \citep[][]{kawa09,perets09}. Indeed this scenario can account 
for many of the observed characteristic of both SNe (namely low explosion energy, very low ejected mass 
and ejection of small amount of $^{56}$Ni), but the possibility to reproduce all the observed 
properties (as the spectroscopic features and, in particular, the alleged
Ca-richness) deserves further investigation.

We are aware that large uncertainties of both theoretical and observational nature are still 
present on the EC-SN mechanism in super-AGB stars. Nevertheless we believe that the scenario 
herein proposed is promising for understanding an increasing number of underenergetic and 
unusual SNe. Only a combined effort will solve the issue. On one side we need more accurate 
observational constraints about the production of intermediate-mass nuclei (specifically C, 
O and all the $\alpha$-elements in general) in low luminosity SN events. On the other side 
more refined future studies on the super-AGB stellar evolution fully describing the TP-SAGB 
phase, and 3-D simulations for examining in detail the nucleosynthesis processes in EC-SNe 
are desirable.

\acknowledgments
M.L.P. acknowledges the support of the {\it ``Padova Citt\`a delle 
Stelle''} prize by the City of Padua. We also thank the referee for 
her/his valuable suggestions to improve our manuscript.

\clearpage

\end{document}